%% file: photon.tex
\documentstyle[12pt,epsfig]{article}

\newcommand{\newc}{\newcommand}

\newc{\Lumi}{{\cal L}}

\newc{\ra}{\rightarrow}
\newc{\Ra}{\Rightarrow}

\newc{\pom}  {I\hspace{-0.2em}P}

\newc{\ftwog} {F_{2}^{\gamma}}
\newc{\ftwop} {F_{2}^{\rm proton}}
\newc{\ftwoc} {F_{2}^{\rm charm}}

\newc{\gev}{\,GeV}

\newc{\rpv}{{\not \!\! R_p}}
\newc{\rpvm}{{\not  R_p}}

\newc{\gsim}{{\stackrel{>}{\sim}}}
\newc{\lsim}{{\stackrel{<}{\sim}}}
\newc{\sleq} {\raisebox{-.6ex}{${\textstyle\stackrel{<}{\sim}}$}}
\newc{\sgeq} {\raisebox{-.6ex}{${\textstyle\stackrel{>}{\sim}}$}}

\def\3{\ss}

\newc{\ETJ}{E^{{\rm Jet}}_T}

\def\ET{E_{\rm t}}

\def\xgo{x_\gamma^{\rm OBS}}

\def\ETAJ{\eta^{jet}}

\def\EPJ{Eur. Phys. J.}
\def\CPC{Comp. Phys. Comm.}
\def\ZFP{Z. Phys.}
\def\photon99{Proceedings of Photon 99, Freiburg, May 1999}

\def\q2{{\rm Q}^2}
\def\p2{{\rm P}^2}
\def\gev2{{\rm GeV}^{2}}

\def\ds{D^{\ast}}
\def\d0{D^{0}}

\def\begr{\begin{flushright}}
\def\endr{\end{flushright}}

\def\begl{\begin{flushleft}}
\def\endl{\end{flushleft}}

\input LP99macros.tex

\def\Title#1{\begin{center} {\Large {\bf #1} } \end{center}}

\begin{document}

\Title{Structure of the Photon}

\bigskip\bigskip

%+\addtocontents{toc}{{\it A.B. Author}}
%+\label{authorStart}

\begin{raggedright}  

{\it J.M. Butterworth\index{Butterworth, J.M.}\\
Department of Physics and Astronomy\\
University College London
London, WC1E 6BT}
\bigskip\bigskip
\end{raggedright}

\section{Introduction}

\subsection{The Structure of a Fundamental Gauge Boson?}

The photon is of course the gauge boson of QED, and is as far as we
know elementary. When we measure photon ``structure'', what we are in
fact doing is probing the quantum fluctuations of the field theory.

The photon couples - via a splitting into virtual, charged
fermion-antifermion pairs - into the electroweak and strong
interactions\index{strong interaction}.  Such behaviour is an
important aspect of quantum field theories, and similar phenomena
arise in many different situations. For example, the gluon splitting
to quarks drives the scaling violations in hadronic structure. Studies
of photon structure test our understanding of this behaviour. In this
talk I give an overview of the current data and phenomenology, and
outline some opportunities available in the medium term future.

\subsection{How photon structure is measured}

The quick answer to the question ``How do you measure the structure of
the photon'' is unfortunately ``With great difficulty''. Experiments
measuring the photon structure in general use the almost on-shell
photons accompanying $e^+$ or $e^-$ beams. These photons are typically
probed by some short distance process. This may be deep inelastic
scattering \cite{epem}, high transverse energy ($\ET$)
jets\index{jets}\cite{jets,zeusxgo,h1effpdf,zeusdijets,zeus3jet} or
particles or heavy quark production\cite{charmpp,xgocharm}.

These are in general rather challenging measurements. One key problem
is the fact that the target photons have a spectrum of energies. If
this is integrated over, sensitivity to the photon structure is
lost. Thus, some way must be found to measure the photon energy on an
event-by-event basis. Another area of difficulty is that although the
presence of high $\ET$ particles does imply the presence of a short
distance scale, the exact relationship between the distance scale and
$\ET$ is not clear.

\begin{figure}[htb]
\begin{center}
\epsfig{file=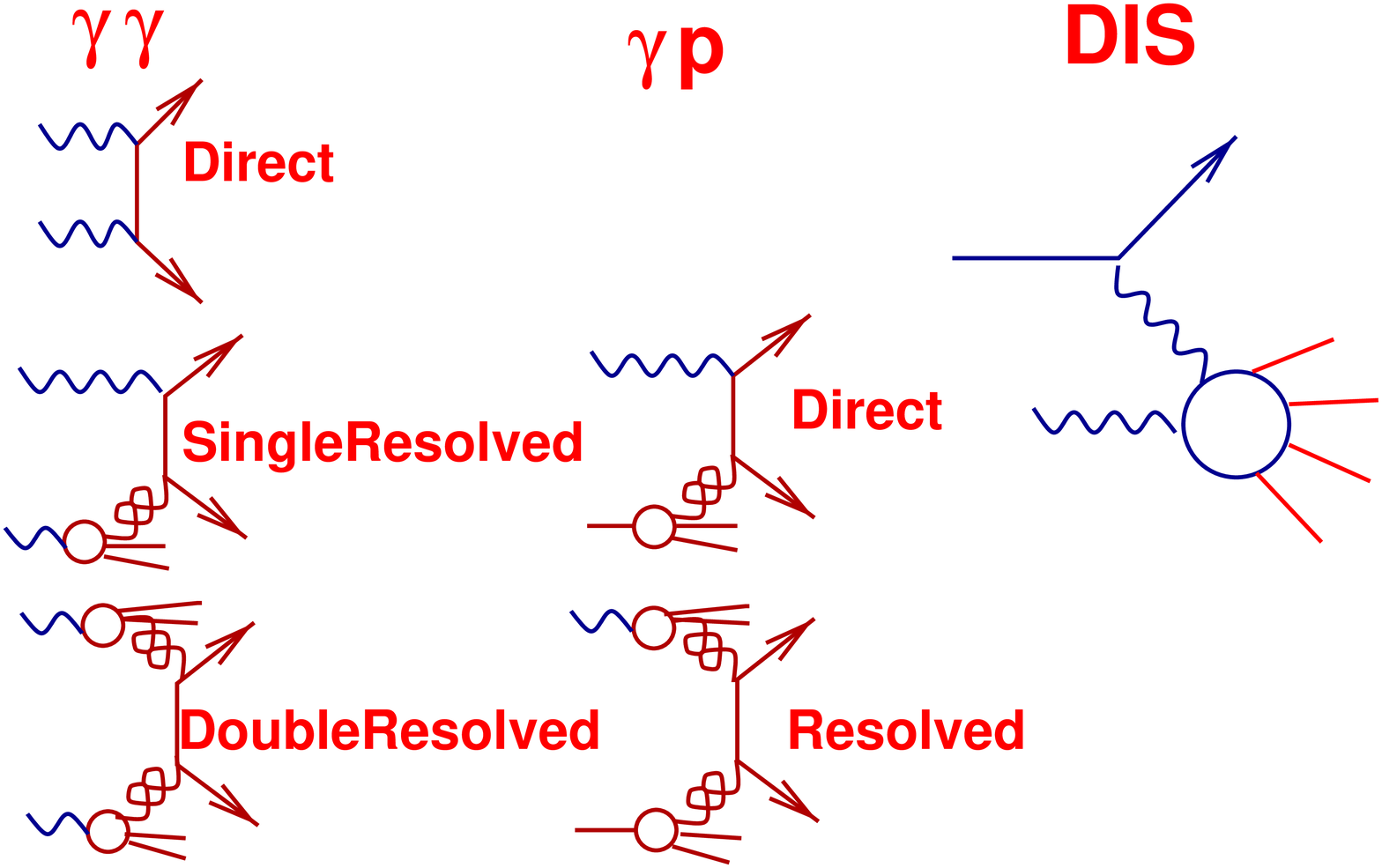,height=5.5cm}
\caption{Examples of leading order diagrams for jets and heavy flavour
production at HERA and LEP, and DIS at $e^+e^-$ machines.}
\label{fig:diags}
\end{center}
\end{figure}

The leading order processes for jet, particle and heavy quark
production are shown in figure~\ref{fig:diags}.  In each case the
virtual parton propagator probes photon at a scale related to $\ET^2$.

The diagrams in which the photon enters into the hard process directly
(``direct processes'') and those in which a partonic photon structure
is resolved (``resolved processes'') have comparable cross sections.
At higher orders, and in real data, separation between resolved and
direct is at some level arbitrary, since if the photon splits to a
$q\bar{q}$ pair of virtuality $\mu$, the process could be categorized
as either resolved or NLO direct, depending upon whether the
factorization scale is chosen to be greater or less than $\mu$.
Despite this, it is still possible and useful to select events in
which a greater or lesser fraction of the photon's momentum enters
into the hard process, where the hard process is defined in terms of
observables such as jets. Such a selection is often made on the basis
of the variable 
\beq 
\xgo = \frac{\sum \ETJ e^{-\ETAJ}}{2yE_e} =
\frac{\sum_{jets} E-p_z}{(E-p_z)_\gamma} 
\eeqn 
which is the fraction the photon momentum entering into the
jets~\cite{zeusxgo}.  Direct processes have high $\xgo$ and resolved
processes have low $\xgo$. Since $\xgo$ is a kinematic variable
defined in terms of jets, it is calculable to any order in QCD and in
any model desired.

The other major process by which photon structure is measured is deep
inelastic $e \gamma$ scattering (also shown in figure~\ref{fig:diags}).
In this case an inclusive measurement of the cross section is made as
a function of the four-momentum transfer at the lepton vertex ($Q^2$)
and/or the Bjorken scaling variables $x$ and $y$.
This allows the extraction of structure functions, defined in exactly
the same way as for a nucleon. Neglecting weak interactions (since for
current experiments $\q2 \ll M_W^2$), this gives: 
\beq
\frac{d^2\sigma_{e\gamma \ra e{X}}}{dxd\q2} = \frac{2\pi
\alpha^2}{xQ^4}\left[ (1+(1-y)^2) {F^\gamma_2 (x,\q2,\p2)} - y^2
{F_L^\gamma (x,\q2,\p2)}\right]
\eeqn 
where if $X = \mu^+\mu^-$ the QED structure is being probed (\ie the
number of muons `in' the photon) and if ${X} = $ hadrons the QCD
structure is being probed (\ie the number of quarks `in' the photon).
For DIS, $\q2 \gg \p2$, that is a ``highly virtual'' photon probes a
less virtual photon at scale $\q2$.

\section{QED Structure Function}

\begin{figure}[htb]
\begin{center}
\epsfig{file=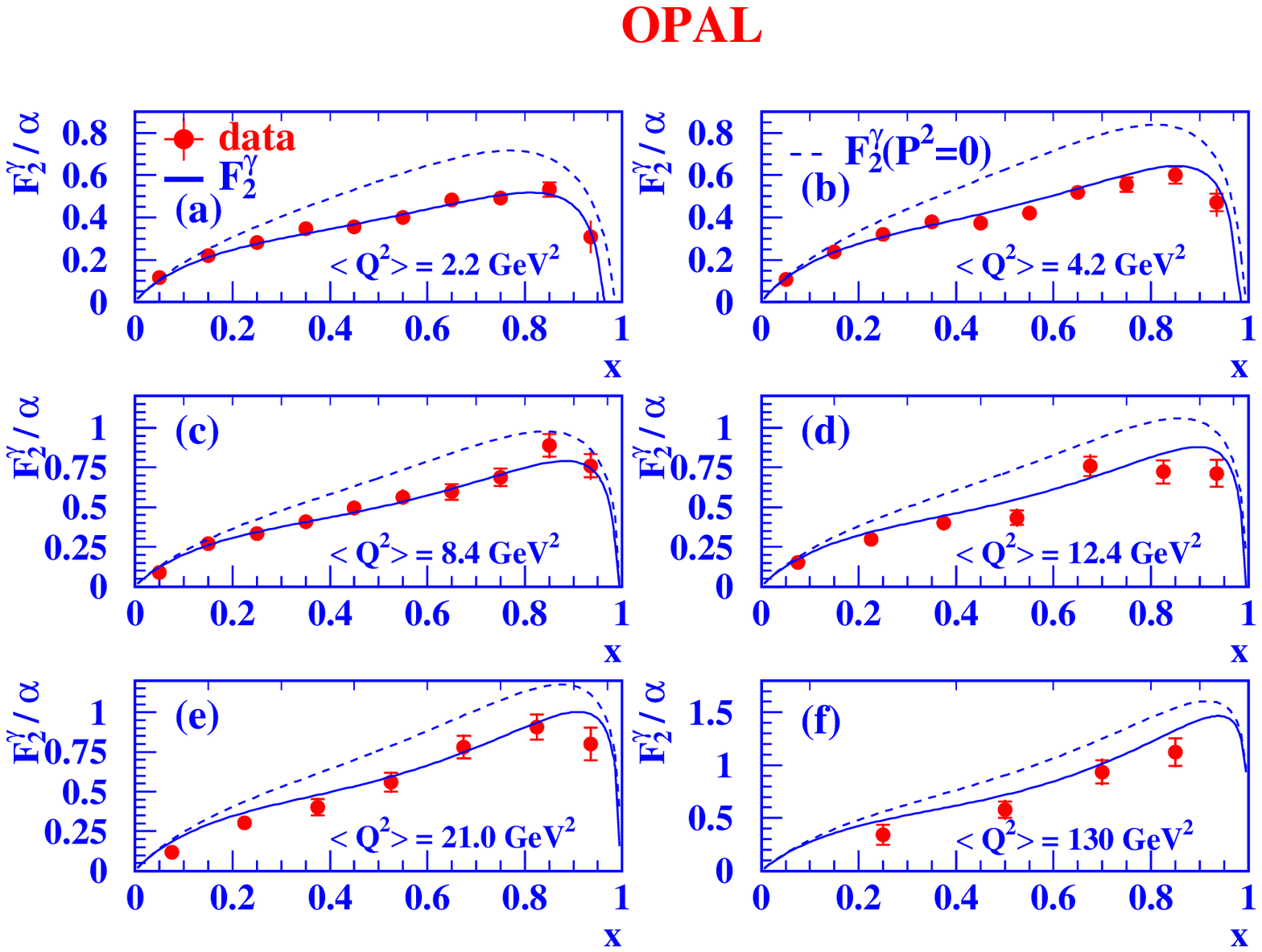,height=7.0cm}
\caption{The QED structure function of the $F_2$ photon. The structure
function is measured by the OPAL Collab. as a function of the Bjorken
scaling variable $x$ in six different bins of the hard scale,
$\q2$. Measurements have also been made by the other LEP
collaborations~\cite{qedsf}.}
\label{fig:QED}
\end{center}
\end{figure}

The QED structure of the photon is exactly calculable, since no strong
interaction is involved. The OPAL measurement is shown in
figure~\ref{fig:QED}. The data are in good agreement with fundamental
prediction of QED (solid line). Measurements have also been made by
the other LEP experiments~\cite{qedsf}. Noteworthy features include
the peak at high $x$ values, and the fact that the structure function
increases with $\q2$.  Both effects are due to the first splitting of
the photon into $\mu^+\mu^-$, a least one of which carries a large
fraction of the photon energy. As $\q2$ increases, smaller and smaller
splittings of the photon can be resolved and thus the structure
function increases, even at high $x$.

Before moving on to compare with the QCD structure, two other points
should be made here. Firstly, the $x$ resolution is fairly good
(around 0.03). This is because $W$, the $\mu^+\mu^-$ mass, is well
measured, and hence the target photon energy is well known.  A second
point is that the virtuality of the target photon has a significant
effect for all $\q2$, even though $\langle \p2\rangle =
0.05~\gev2$. This can be seen by comparing the solid line, where the
virtuality has been taken into account, with the dashed line, where
the target photon has been assumed to be real.

\section{QCD and the `Real' Photon}

The case in which the photon splits into a $q\bar{q}$ pair is
theoretically more complex. If the quarks have a low transverse
momentum ($k_T$) with respect to each other, they can exist for times
which are long on the time-scale of the the strong interaction.  Thus,
a complex partonic system can evolve, consisting of a mixture of
perturbative and non-perturbative physics.  In addition to the
presence of non-perturbative physics in the initial state, the final
state also involves long-distance hadronization processes.

The QCD structure of the photon is also experimentally more
problematic.  Jets of hadrons are in general harder to measure than
muons.  This does not just affect measurements of jet or particle
production - where issues such as fragmentation and `underlying
events'\index{underlying event} become important. It is also an issue
in DIS, since the photon energy must be measured from the final
state. Thus it becomes critical to contain as much of the hadronic
system as possible in the detector, and vital to have the best
possible estimates of the structure of this final state to allow
estimates of the resolution and acceptance to be reliably
made~\cite{photon95}.

These effects mean that there is potentially large model dependence in
any extraction of photon structure. Dealing with this model dependence
is a {\it major} issue for the experiments. Best practice is to
separate the measurement as much as possible from interpretation. Thus
measurements of cross sections are made with the minimal model
dependence. These cross sections are defined in terms of the real
final state (hadrons rather than partons!) and the kinematic regions
in which they are measured are dictated by detector
acceptance. Because of this, they are unfortunately often hard to
interpret and compare with each other or with fundamental QCD.

These cross sections are sensitive to many important effects, not only
the photon structure, but that of the proton (in particular the gluons
at $x \approx 10^{-2}$ at HERA), as well as $\alpha_s$, low-$x$ QCD,
hadronization and ``underlying'' events\index{underlying event}. There
is not much benefit in being sensitive to all these at the same time,
and thus the role of phenomenology is critical to isolate as far as
possible different effects. Several programs allowing flexible
calculation of the processes in QCD at NLO are
available~\cite{nlofr,nloprogs} and are crucial in the ongoing effort to
extract fundamental physics from a wide range of data. Equally
important are general purpose Monte Carlo
simulations~\cite{mcs} which include models of
hadronization and underlying events as well as (typically) LO QCD and
parton showers. These allow us to build a consistent picture of these
processes over a large data set from several experiments, and also
allow improvements in our understanding to be fed back into modelling
of detector acceptance, leading to improved measurements.

\begin{figure}[htb]
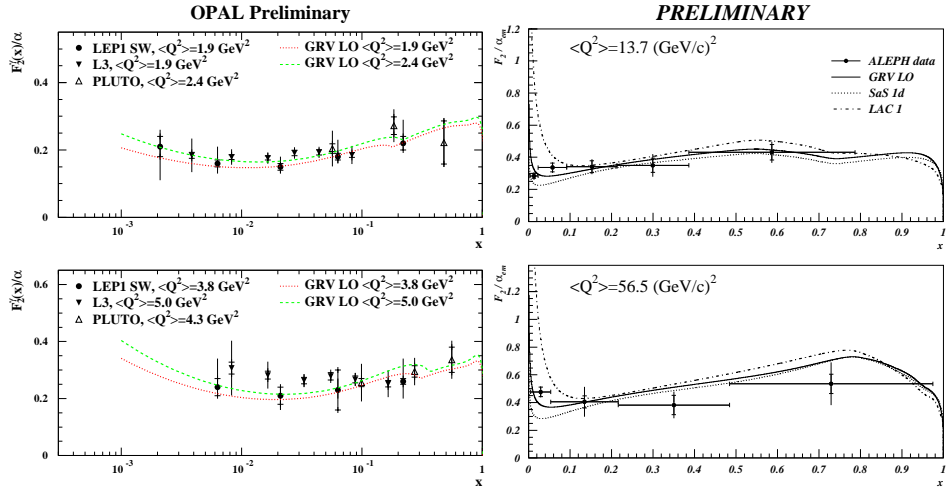

\begin{center}
\epsfig{file=result1others.epsi,height=6.4cm,clip=}
\epsfig{file=f2_result.epsi,height=6.4cm,clip=}
\caption{Examples of recent data on the QCD structure function $F_2$
of the photon (see text).}
\label{fig:f2g}
\end{center}
\end{figure}

The potential rewards are very high, and these are exciting times for
those involved. However, it is very much ``work in progress'', and the
following represents a snapshot of an evolving field. See
these~\cite{workshops} references for more details.

\subsection{QCD Structure Function}

The QCD structure of the photon has been measured for samples with
$0.24~\gev2 < \langle \q2\rangle < 400~\gev2$~\cite{epem}. An example
of the more recent data at low $x$ and intermediate $\q2$ is shown in
figure~\ref{fig:f2g}a, from OPAL and L3. Earlier PLUTO data are also
shown. The data are compared to curves from the GRV
group~\cite{grv}. In the GRV model of the proton, the rise in $\ftwop$
is generated by the DGLAP evolution~\cite{dglap}. The same behaviour
is expected in the photon parton distributions at similar $x$, and can
be seen in the curves. Unfortunately the data do not yet have the
reach or accuracy to determine whether or not such a rise
occurs. However, improved statistics and smaller systematic
uncertainties are expected before the end of LEP running. It should be
noted that L3 sometimes display their data without showing an estimate
of the model dependence, preferring instead to show a series of data
sets each corrected according to a different model. Here, these
different data sets have been used to estimate the systematic
uncertainty~\cite{nisius}.

Figure~\ref{fig:f2g}b shows preliminary measurements at a higher $\q2$
from ALEPH. Since these are plotted on a linear scale in $x$, they can
easily be compared to the QED structure functions of
figure~\ref{fig:QED}. Whilst at low $x$ the gluon splitting means that
the structure function is expected to rise like that of the proton, at
higher $x$ the photon to two fermion splitting dominates and the
behaviour is similar for the QED and QCD structure functions.

\begin{figure}[htb]
\begin{center}
\epsfig{file=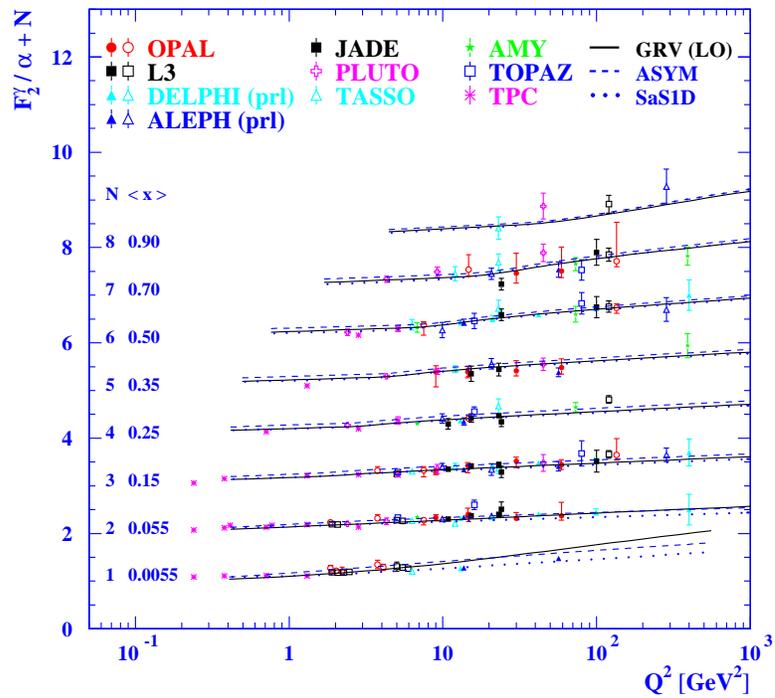,height=9.5cm,clip=}
\caption{$\q2$ dependence of the QCD structure function.}
\label{fig:q2dep}
\end{center}
\end{figure}

Also clearly seen in the data is the expected positive scaling
violation at all $x$, driven by photon splitting at high $x$ and (like
the proton) by the gluons at low $x$. The data are summarized in
figure~\ref{fig:q2dep}.

\subsection{QCD and the Real Photon at HERA}

Since at HERA the photon is probed by a parton from the proton, HERA
does not measure $\ftwog$.  The HERA equivalent of $\ftwog$ is a jet
cross section. This has the major disadvantage that hadronization, as
well as choice of jet definition, plays a role. An advantage, however,
is that the gluon distribution in the photon enters directly in the
cross section at leading order. A further advantage is that due to the
fact that the \CM~frame is boosted strongly in the proton direction,
the photon remnant tends to open out and be relatively well
measured in the detector. In addition, both ZEUS and H1 have small
angle taggers, which allow the photon energy to be inferred from
the electron energy. These effects mean that the target photon energy
is better measured than at LEP.

The measured cross sections may be compared to NLO pQCD calculations,
which take a photon parton distribution function (PDF) as input. If
the jets have high enough transverse energy ($\ETJ$) the hadronization
corrections are expected to be at the level of a few percent. The
probing scale is something of the order of $\ETJ$.

\begin{figure}[htb]
\begin{center}
\epsfig{file=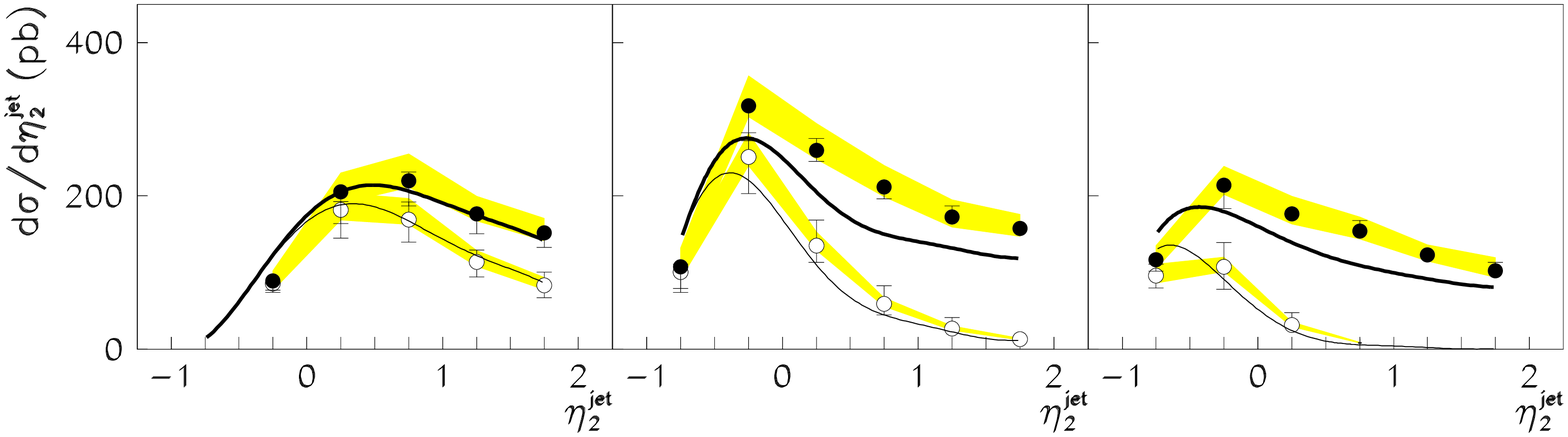,width=7.0 cm}
\epsfig{file=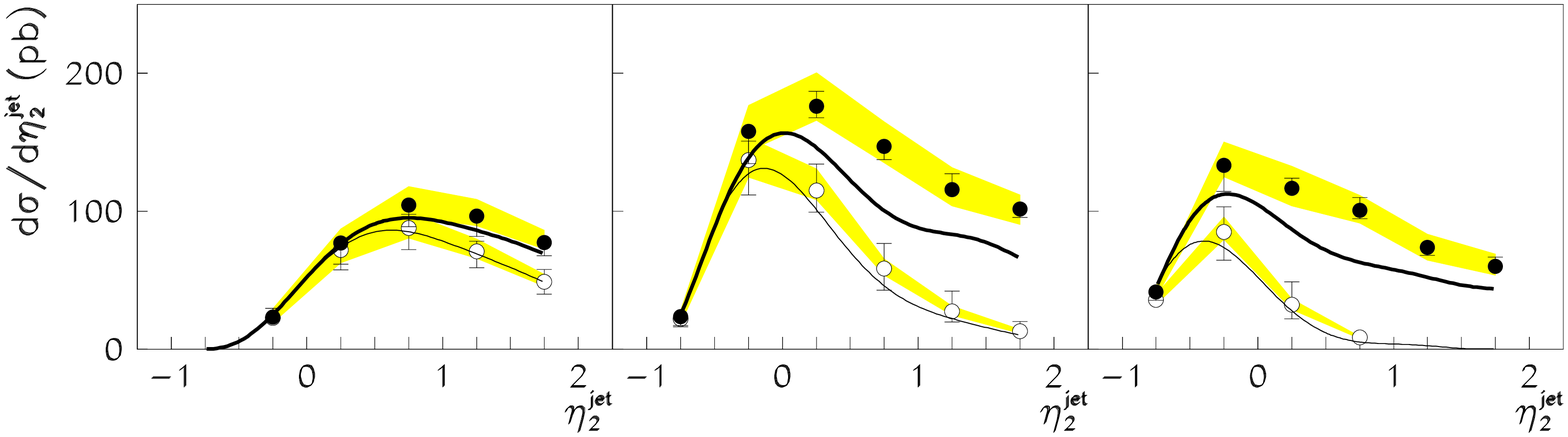,width=7.0 cm}
\epsfig{file=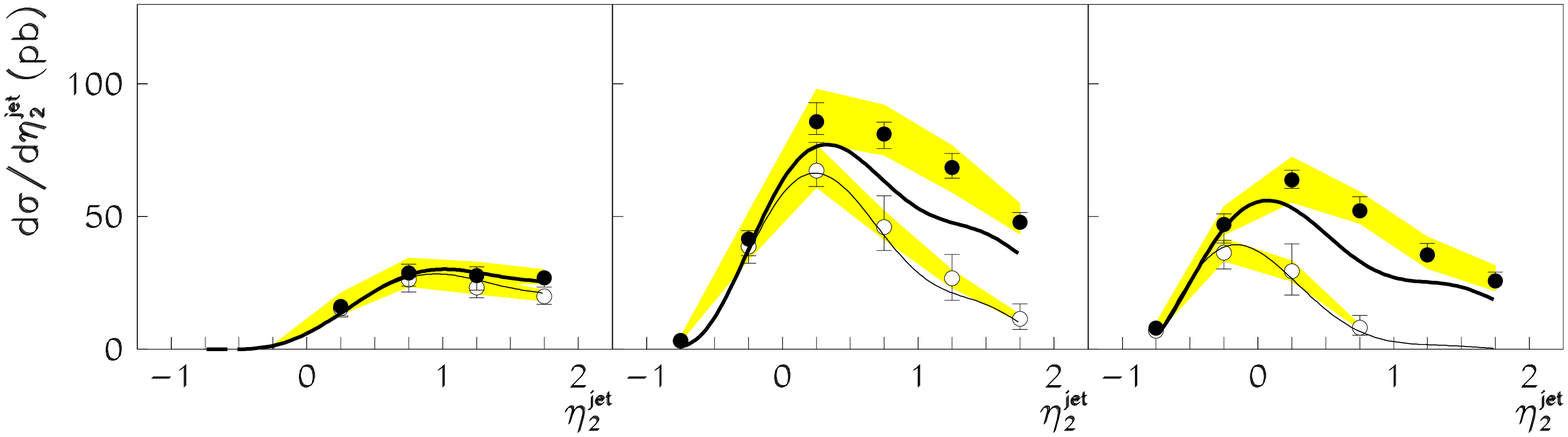,width=7.0 cm}
\epsfig{file=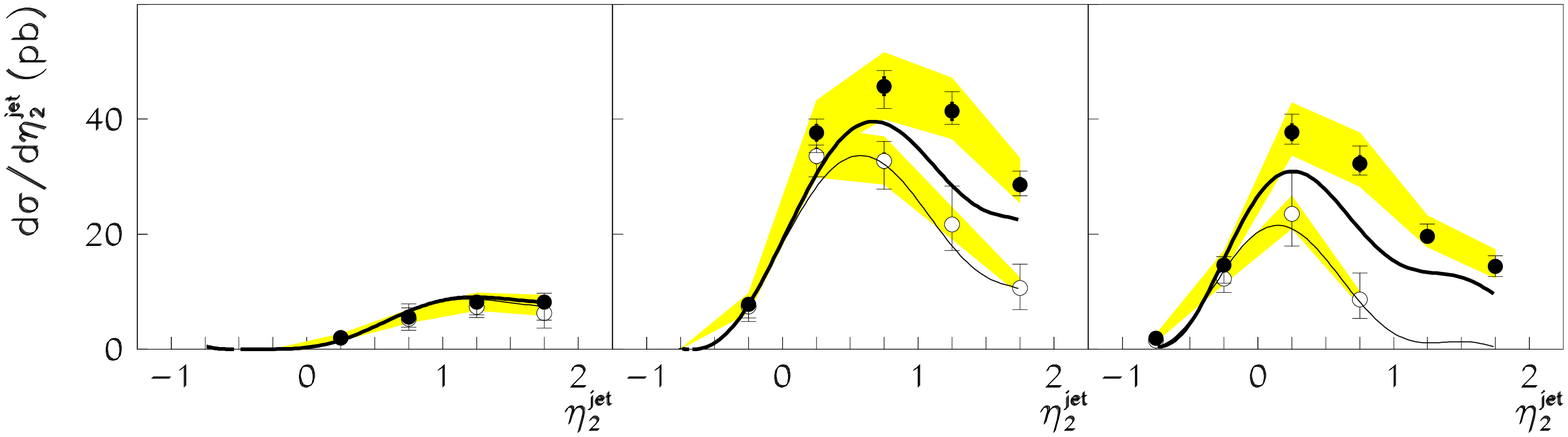,width=7.0 cm}
\epsfig{file=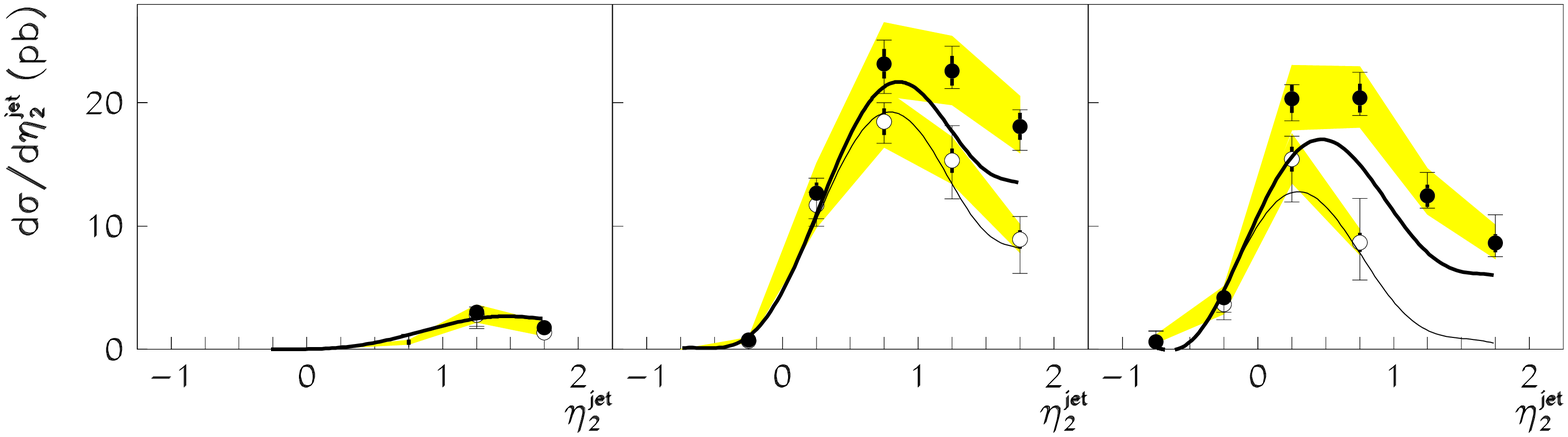,width=7.0 cm}
\caption{Dijet cross sections at HERA.}
\label{fig:dijets}
\end{center}
\end{figure}

The latest ZEUS preliminary data are shown in figure~\ref{fig:dijets}
for differential cross sections defined as in~\cite{zeusdijets} but
now measured above a variety of $\ETJ$ thresholds, increasing the hard
scale. The data are compared to a calculation~\cite{nlofr} using the
AFG-HO Photon PDF~\cite{afg}.  The high $\xgo$ data is in excellent
agreement with the theory. However, in the region including both high
and low $\xgo$ data, there is a discrepancy particularly in the
forward region, where the lowest values of $\xgo$ are probed. This
discrepancy shows no sign of dying away with increasing $\ETJ$ even
though hadronization effects are estimated to be small at these
values.

\begin{figure}[htb]
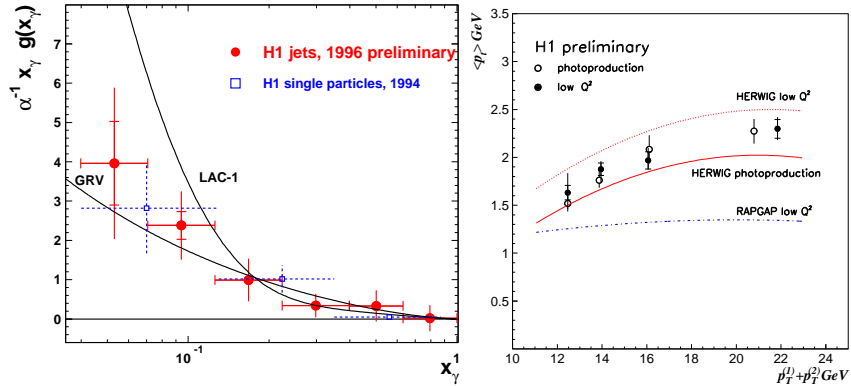

\begin{center}
\epsfig{file=gluonoldstat.h1all.epsi,width=6.0cm}
\epsfig{file=mean_PT_H1prel.epsi,height=5.0cm}
\caption{(a) Effective photon PDF~\cite{h1effpdf}, (b) Transverse
energy of the photon~\cite{h1rem} remnant.}
\label{fig:effrem}
\end{center}
\end{figure}

The potential of such data may be illustrated by assuming LO QCD \& MC
models, and estimating a ``parton level cross section'', from which an
effective parton density can be extracted. This exercise has been
performed by H1, both in the case of jets and charged particles and
the result is shown in figure~\ref{fig:effrem}a. The rise in the gluon
distribution, which will drive the rise in $\ftwog$ at lower $x$, is
clearly in the region of HERA sensitivity, although the model
dependence in the data is now large.

The physics of hard scattering in photon-proton collisions is far from
trivial. The photon remnant is an interesting feature of the final
state. Some of its properties have been measured by
ZEUS~\cite{zeusrem}. In particular it was measured to have an average
transverse momentum $p_T = 2.1 \pm 0.2$~GeV w.r.t. photon direction.
These measurement have now been extended by H1, measuring the remnant
as a function of $\ETJ$ for photoproduction and for virtual photons
($1.4 < \p2 < 25~\gev2$). These results, shown in
figure~\ref{fig:effrem}b, are consistent ZEUS result.  Importantly,
the behaviour of the photon remnant is critical for the $x$ resolution
at LEP, since it determines how much hadronic energy escapes down the
beam-pipe. HERWIG (shown in the figure) does a reasonable job, and
such distributions are used to constrain the models employed at LEP,
thus reducing the systematic errors.

\begin{figure}[htb]
\begin{center}
\epsfig{file=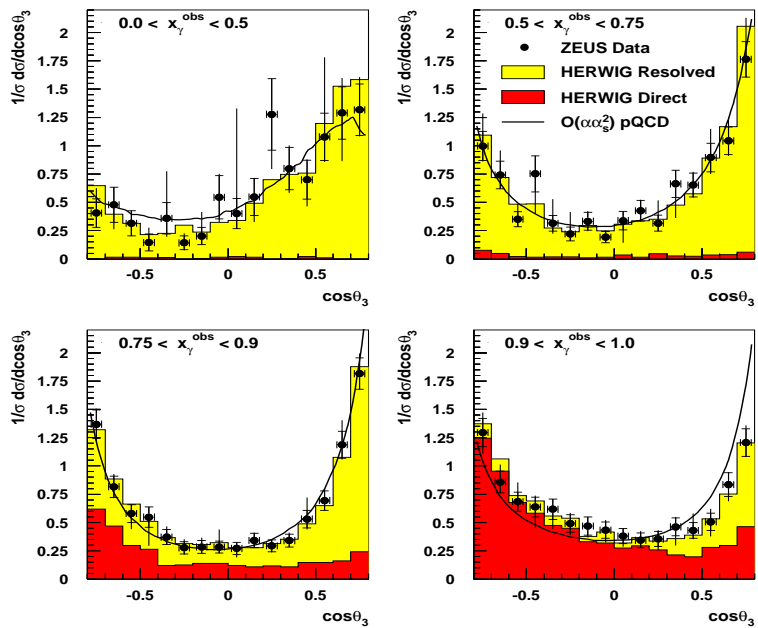,width=10cm,height=9cm}
\caption{Three jet cross sections at HERA. LO resolved photon
processes are shown as red, and LO direct as yellow.}
\label{fig:3jet}
\end{center}
\end{figure}

The fact that the photon has a dual nature - behaving either as a
hadron or a point-like particle - allows several interesting QCD
studies to be made.  A recent measurement is that of the three-jet
distributions. The QCD dynamics of the three jet system is sensitive
to the colour of the incoming partons. In figure~\ref{fig:3jet},
$\theta_3$, the angle between the highest energy jet and the proton beam
direction (defined as in~\cite{zeus3jet}), is shown, and compared to
${\cal O}(\alpha\alpha_s^2)$ QCD and to LO MC simulation, in which the
third jet comes from the parton shower.  There is a change in shape of
the distribution as $\xgo$ increases and the mix of incoming resolved
and direct photons changes.

\begin{figure}[htb]
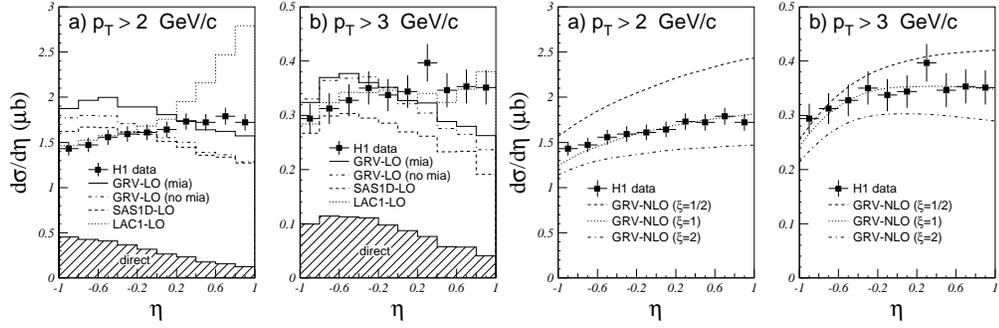

\begin{center}
\epsfig{file=neweta4.epsi,width=6.5cm}
\epsfig{file=neweta3.epsi,width=6.5cm}
\caption{Charged particle distributions~\cite{h1effpdf}. The curves
marked ``mia'' or ``no mia'' have multiparton interactions turned on
or off respectively.}
\label{fig:particles}
\end{center}
\end{figure}

Charged particle distributions are also sensitive to photon PDFs. They
also require non-perturbative input in the form of a fragmentation
function, but once this is taken into account, there are no
hadronization uncertainties as such. However, there is still
sensitivity to the modelling of the underlying event\index{underlying
event} and the choice of hard scale (see figure~\ref{fig:particles}).

In addition to these processes there are prompt photon
data~\cite{promptp}, as well as measurements of jet shapes and
sub-jets at HERA and LEP~\cite{shapes}, all of which have the power to
reduce the uncertainties in the final state and in the theory, if
taken together. Given this enormous data set, of increasing accuracy
and scope, {\it the time is right to do a serious QCD fit to the HERA
and LEP data!}

In the final sections of this presentation is describe briefly two
areas of photon structure studies. Both are relatively new, and both
offer new and possibly simpler ways to investigate the underlying
physics.

\section{Charm and the Real Photon}

\begin{figure}[htb]
\begin{center}
\epsfig{file=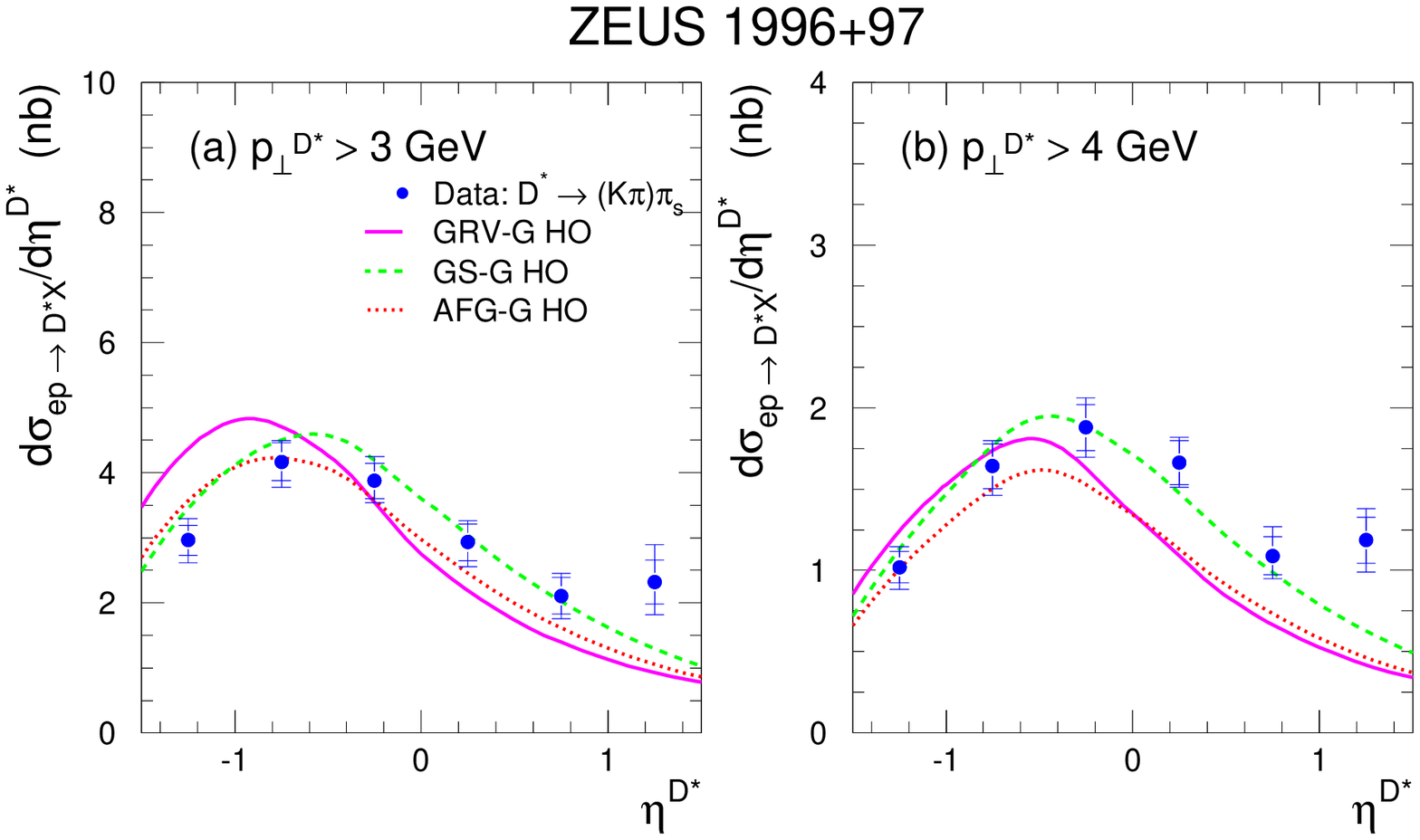,width=8.5cm}
\epsfig{file=pn390_06.epsi,width=4.75cm}
\caption{Inclusive $\ds$ photoproduction cross sections from HERA and
LEP~\cite{charmpp}. The data are compared to~\cite{charmtheory}}
\label{fig:incstar}
\end{center}
\end{figure}

Charm photoproduction\index{charm photoproduction} has been measured
at both HERA and LEP~\cite{charmpp,xgocharm}. If it is assumed that
the charm mass is sufficiently high that perturbative QCD is
applicable, then the `charm content' of the photon is expected to be a
totally perturbatively calculable parton distribution. In fact if the
factorization scale is taken lower than the charm mass, charm
production takes place entirely within the hard process and there is
no charm content to the photon. The data are beginning to address
whether this picture is sufficient. Charm is typically tagged in the
$\ds$ channel and recent measurements of $\ds$ cross sections are
shown in figure~\ref{fig:incstar}.

\begin{figure}[htb]
\begin{center}
\epsfig{file=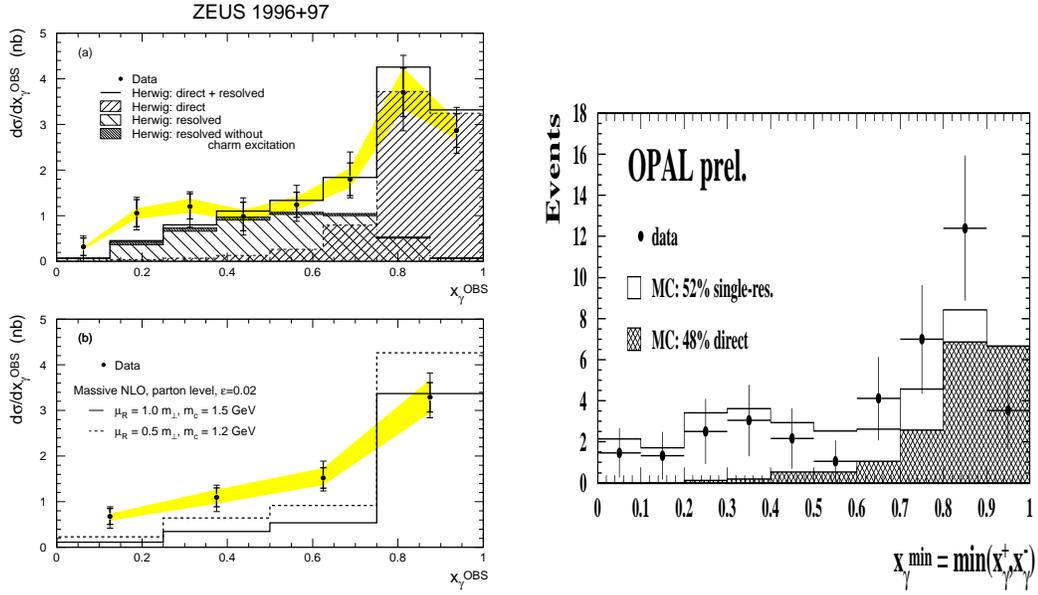,width=7.0cm}
\epsfig{file=pn390_03.epsi,width=6.5cm,height=6.5cm}
\caption{ZEUS cross section and OPAL event distribution as a function
of $\xgo$ from charm-tagged dijet events~\cite{xgocharm}.}
\label{fig:xgocharm}
\end{center}
\end{figure}

The agreement with theory is reasonable. However, the theory lies
somewhat below the data in the forward (proton) direction in the HERA
data. If jets are measured, $\xgo$ can be calculated and we can begin
to examine the production mechanism in more detail. The $\xgo$
distributions from ZEUS and OPAL are shown in
figure~\ref{fig:xgocharm} where at least one jet contains a $\ds$.

Comparison to the LO Monte Carlo shows that direct and resolved
processes are both needed. This is still true even when the data is
compared to NLO QCD~\cite{nlofr}. The photon PDF used in the
calculation contains no charm, but events can be generated at low
$\xgo$ where a third jet plays the role of the photon remnant.  The
resolved processes are suppressed relative to direct in comparison to
the non charm-tagged case, but the cross section is still significant
at low $\xgo$.  Beauty in photoproduction has also been seen now at
both HERA and LEP~\cite{beauty}.

\section{Virtual Photon Structure?}

All current studies of the structure of the photon in fact study
photons with a finite (if small) virtuality. Nevertheless, there has
been a marked discontinuity in the terminology and methodology used to
describe the photon as a propagator (in electron-proton DIS for
example) and the photon as a target.  This is despite the fact the $ep$
``DIS'' experiments extend well down below 1~GeV in $\q2$, where the
term ``deep'' inelastic scattering is arguably inappropriate, and
also despite the fact that, as seen in the QED structure results
(figure~\ref{fig:QED}), the effect of target photon virtualities is
significant even in so-called ``photoproduction'' experiments.

This situation is changing, and a significant amount of attention is
now being paid both by theory and experiment to the fact that there
must be a continuum between $\p2 = 0$ and $\p2 \approx \q2$. There
exist some expectations as to how the transition between the two might
take place. With respect to direct photon processes, the expectation
is that the perturbative part of the photon structure will fall like
$\ln(\q2/\p2)$ whilst the non-perturbative (``Vector Meson'') part
should fall something like $m_v^2/(m_v^2 + \p2)$, where $m_v$ is a the
mass of the vector meson state into which the photon may fluctuate.
When the photon virtuality gets large, but remains much less than the
probing scale (which may be set by high $\ETJ$ jets, for example), the
non-perturbative part vanishes and once more we obtain a
perturbatively calculable parton distribution, which suggests
possibilities for the measurement of $\alpha_s$ and an improved
understanding of QCD radiation and hadronic structure.

\begin{figure}[htb]
\begin{center}
\epsfig{file=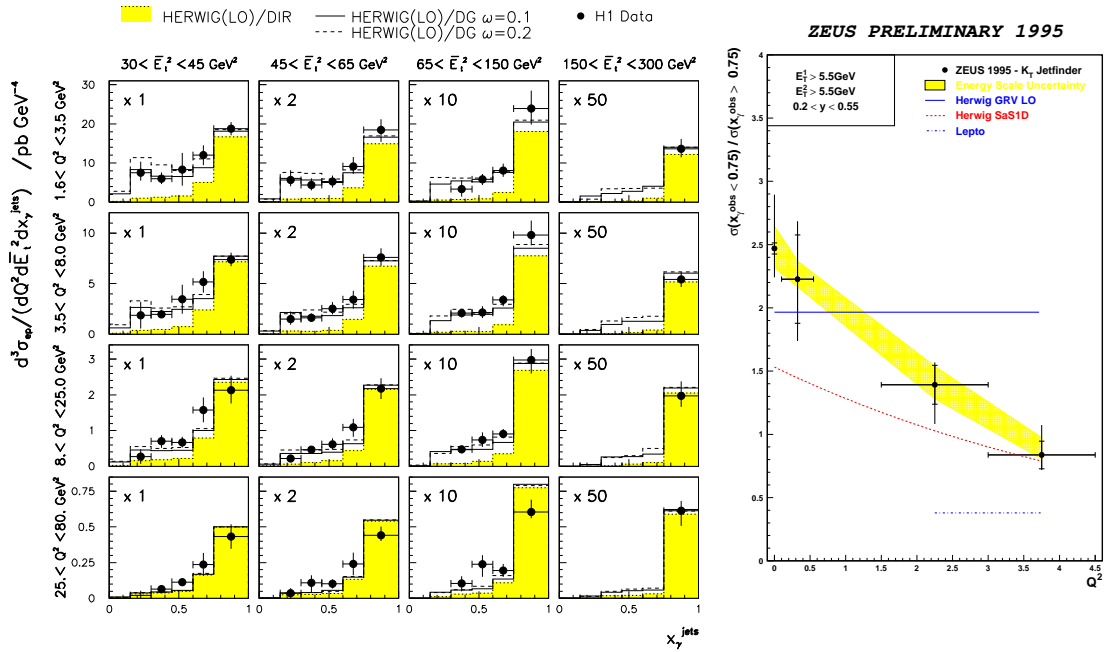,width=9.5cm}
\epsfig{file=symm_ratio.epsi,width=5cm}
\caption{Virtual photons resolved.}
\label{fig:vphera}
\end{center}
\end{figure}

Look first at the HERA data, some of which is shown in
figure~\ref{fig:vphera}~\cite{heravp}.  In figure~\ref{fig:vphera}a,
H1 dijet data $d\sigma/d\xgo$ is shown in a grid in which the probing
scale ($\approx (\ETJ)^2$) increases from left to right, whilst the
target scale (the photon virtuality $\p2$, here labelled $\q2$
according to the convention at HERA). Concentrate for instance on the
second row. The target photon virtuality is in the range $3.5~\gev2 <
\p2 < 8.0~\gev2$, certainly far from zero, and yet the population of
events at low $\xgo$ is significant. The LO QCD plus parton showers
simulation can only successfully model the distribution by appealing
to a virtual photon structure ansatz~\cite{dgsupp} in which the
expectations above are implemented. A similar effect is observed in
the ZEUS measurement (figure~\ref{fig:vphera}b) where the ratio of the
high to low $\xgo$ cross sections is plotted, now extending all the
way down to the ``almost real'' photons previously studied. The ratio
falls rapidly. However, at the lowest target virtualities it is higher
than the expectations shown, and even by $4~\gev2$ it remains higher
than the expectation of a DIS Monte Carlo which contains no photon
structure. The blue line is a ``straw man'' model in which the GRV
real photon PDF has been used in virtual photons without
modification. It is not expected to be valid here, but the fact that
it is completely flat demonstrates graphically that the observed fall
in the data is genuinely due to suppression of the photon structure,
and not to any subtle phase space effect. The Schuler and Sj\"ostrand
parton distribution function (red curve) contains a model for the
virtual photon structure which a suppression with increasing
virtuality.  It is interesting to note that whilst both curves lie
below the data at the lowest virtuality, the SaS prediction falls more
slowly than the data and thus there is agreement at the higher
virtualities. The discrepancy at low virtuality (and at jets $\ETJ$ of
around 6~GeV, where these data lie) has been observed before and
attributed to the effect of a so-called ``underlying event'', possibly
generated by multiparton interactions\index{multiparton
interactions}. Such effects are not included in the curves shown
here. Since models of underlying events rely upon the hadronic nature
of the photon, it is natural that any discrepancy due to them should
fall as the hadronic component is suppressed.

%\begin{figure}[htb]
%\begin{center}
%\epsfig{file=combo.eps,width=6.5cm}
%\caption{Virtual photon cross sections in $e^+e^-$.}
%\label{fig:vplep}
%\end{center}
%\end{figure}

%In figure~\ref{fig:vplep}, some examples of the $e^+e^-$ data on
%virtual photons are shown~\cite{eevp}. Both the early PLUTO data and
%the more recent L3 data (as well as recent OPAL data, not shown) are
%consistent with being flat with $\p2$, but are also consistent with
%the expected fall.

Measurements of virtual photon structure have also been made in
$e^+e^-$ experiments~\cite{eevp}. Both the early PLUTO data and the
more recent LEP data are consistent with being flat with $\p2$, but
are also consistent with the expected fall.

There is also a relation between virtual photon structure and low-$x$
physics: Two virtual photons in collision is as near as we are likely
to get to a ``golden'' process in which the total cross section is
expected to be governed by a pomeron (multi-gluon colour-singlet
exchange) calculable in perturbative QCD according the the
BFKL~\cite{bfkl} resummation. Such processes have been measured at
LEP.  Both leptons are tagged and so there is a good measurement of
both photon virtualities. In contrast to the previous situations,
these virtualities are now selected to be of comparable size.  The
idea is that there should be a large evolution in $x$ (actually in
rapidity), and that the high virtualities mean that non-perturbative
effects should be small. These are the conditions in which the BFKL
resummation of $\ln (1/x)$ terms should be applicable.  Although the
measurements so far are above the naive two-gluon exchange
calculation, they are also consistent with the model encoded in
PHOJET. There is a large uncertainty in the actual prediction of BFKL,
and a conclusive test has yet to be made. This field is developing
rapidly in both theory and experiment.

\section{Summary}

This is a field in which lots of new data has appeared over the past
two years, and more is expected soon. The new results from LEP and
HERA demonstrate the improvements being made in understanding of
hadronic initial and final states, using the photon as a flexible test
case. This has been made possible by the emergence of several new
theoretical tools, including better general purpose simulations,
implementations of virtual photon PDFs, and NLO QCD calculations which
allow realistic kinematic cuts to be applied. Such efforts are proving
critical in extracting fundamental physics from the data.

The final word from LEP, LEP2 and pre-upgrade HERA will be a series of
measurements with much reduced systematic uncertainties over a very
wide kinematic range. The data and theoretical tools are now in place
for a comprehensive analysis of photon structure along the lines of
those carried out for the proton. To challenge our ideas about QCD
structure in a second hadron-like object, with the expected
differences and similarities described in this review, is a great
opportunity and promises to set the essential technology of reliable
QCD calculations on a significantly firmer footing.

In the slightly longer term future, charm and beauty photoproduction
will be a boom area at HERA after the upgrade, as both the luminosity
and the ability of the detectors to tag heavy flavours should increase
markedly.

I believe that the curious nature of the photon, in which by making
judicious selection we can turn on or off its ``hadronic structure''
is an enormously valuable tool for understanding hadronic initial and
final states in general, a topic of increasing importance across the
breadth of particle physics.

I am very glad to acknowledge to all the hard work involved on the LEP
and HERA experiments, as well as the clearly written papers for EPS
and particularly all the lively discussions with many people at
Photon99  - I'm looking forward to
Photon2000. Extra thanks are due to Richard Nisius for several of the
summary plots.

\end{document}

%% file: LP99macros.tex
\def\beq{\begin{equation}}
\def\eeq#1{\label{#1}\end{equation}}
\def\eeqn{\end{equation}}

\def\beqa{\begin{eqnarray}}
\def\eeqa#1{\label{#1}\end{eqnarray}}
\def\eeqan{\end{eqnarray}}

\def\NPB{ Nucl. Phys. B}
\def\PLB{Phys. Lett.  B}

\def\PRD{Phys. Rev. D}

\def\ZPC{Z. Phys. C}

\let\bar=\overbar

\def\ie{{\it i.e.}}

\def\Dslash{\not{\hbox{\kern-4pt $D$}}}
\def\dslash{\not{\hbox{\kern-2pt $\del$}}}

\def\CM{{\mbox{\scriptsize CM}}}

\def\msb{{\bar{\ssstyle M \kern -1pt S}}}

%% file: photon.bbl
\begin{thebibliography}{99}
%%
%%  bibliographic items can be constructed using the LaTeX format in SPIRES:
%%    see    http://www.slac.stanford.edu/spires/hep/latex.html
%%  SPIRES will also supply the CITATION line information; please include it.
%%

\bibitem{epem}
OPAL Collab., \PLB 411 (1997) 387-401; Z. Phys. C61 (1994) 199;
\ZFP C74 (1997) 33; \PLB 412 (1997) 225, E. Clay \photon99.\\
DELPHI Collaboration, \ZFP 69 (1996) 223.\\
L3 Collaboration, Phys. Lett. {B436} (1998) 403; \PLB 447 
(1999) 147; F. Erne \photon99.\\
TOPAZ Collaboration, \PLB 332 (1994) 477.\\
PLUTO Collaboration,  \PLB 107 (1981) 168; \PLB 142 (1984) 111; \ZPC 26 
(1984) 353; NPB 281 (1987) 365.\\
A.B\"ohrer, ALEPH Collab., \photon99.

\bibitem{jets}
H1 Collab., Z.Phys. C70 (1996) 17-30; \PLB 314 (1993) 436.\\
ZEUS Collab., \EPJ C4 (1998) 591; \PLB 384 (1996) 401-413; \EPJ C1 
(1998) 1/2, 109.\\
OPAL Collab., Eur.Phys.J. C10 (1999) 547-561.

\bibitem{zeusxgo}
ZEUS Collab., \PLB 348 (1995) 665-680.

\bibitem{h1effpdf}
H1 Collab., Eur.Phys.J. C10 (1999) 363-372 and 
Eur.Phys.J. C1 (1998) 97-107.

\bibitem{zeusdijets}
ZEUS Collab., Eur.Phys.J. C11 (1999) 35-50.\\
J. Vossebeld,  hep-ex/9909039, International Europhysics Conference on
HEP, Tampere, Finland, July 15-21, 1999, to appear in the proceedings.

\bibitem{zeus3jet}
ZEUS Collab., \PLB 443 (1998) 394-408.

\bibitem{charmpp}
L3 Collab., hep-ex/9909006.\\
ZEUS Collab., \PLB 401 (1997) 192-206.

\bibitem{xgocharm}
OPAL Collab., hep-ex/9911030, submitted to Eur.Phys.J. C\\
ZEUS Collab., Eur.Phys.J. C6 (1999) 67-83.

\bibitem{qedsf} 
 CELLO Collab., \PLB 126B, 384--390 (1983).\\
 DELPHI Collab., \ZFP C69, 223--234 (1996).\\
 L3 Collab., \PLB 438, 363--378 (1998).\\
 OPAL Collab., hep-ex/9902024\\
 PLUTO Collab., \ZFP C27, 249--256 (1985).\\
 TPC/2$\gamma$ Collab., M.P.~Cain et~al., \PLB 147, 
 232--236 (1984).

\bibitem{photon95} 
J. Forshaw, M. Seymour, Proceedings of Photon 95,
3-9, (Sheffield) World Scientific, Ed: D.Miller, S.Cartwright, V.Khoze.

\bibitem {nlofr} S. Frixione, \NPB 507 (1997) 295.

\bibitem{nloprogs}
B.W. Harris and J.F. Owens, \PRD 56 (1997) 4007. \\
M. Klasen and G. Kramer, \ZPC 76 (1997) 67, 
M. Klasen, T. Kleinwort and G. Kramer, Eur.\ Phys.\ J.\ Direct C1 (1998) 1.\\
B.~P\"otter, Comp.~Phys.~Comm. 119 (1999) 45

\bibitem{mcs}
G. Marchesini et al., \CPC 67 (1992) 465 and hep-hp/9607393.\\
H.U. Bengtsson and T. Sj\"ostrand, \CPC 46 (1987) 43 and
T. Sj\"ostrand, \CPC 82 (1994) 74.\\
R. Engel, \ZPC 66 (1995) 203.

\bibitem{workshops}
Proceedings of the HERA Monte Carlo Workshop 1998/99, DESY, 
Ed: A.T.Doyle, G.Grindhammer, G.Ingelman, H.Jung\\
http://www.desy.de/\~ ~heramc/proceedings/\\
A. Finch et al, J.Phys. G24, (1998) 457.\\
A. Finch (LEP experiments) \photon99.

\bibitem{grv} M. Gl\"uck et al., \PRD 46 (1992) 1973.

\bibitem{dglap}
V.N.~Gribov and L.N.~Lipatov, Sov.~J.~Nucl.~Phys.~15 (1972) 438 and 675;\\ 
Yu.L.~Dokshitzer, Sov.~Phys.~JETP 46 (1977) 641;\\
G.~Altarelli and G.~Parisi, Nucl.~Phys.~B126 (1977) 297.

\bibitem{nisius} R. Nisius, hep-ex/9907012, Invited talk given at 
Photon 99, Freiburg, May 1999.

\bibitem {afg} P. Aurenche et al., \ZFP C64 (1994) 621.

\bibitem{zeusrem}
ZEUS Collab., Phys.Lett. B354 (1995) 163-177.

\bibitem{h1rem} Alice Valkarova (H1 Collab.), \photon99.

\bibitem{promptp}
ZEUS Collab., \PLB 413 (1997) 201-216.

\bibitem{shapes}
ZEUS Collab., Eur.Phys.J. C2 (1998) 61-75.\\
Sabine W\"olfle (ZEUS Collab.) \photon99.

\bibitem{charmtheory}
B.A.Kniehl et al.,
  \ZPC 76 (1997) 689; \\
J.Binnewies et al., hep-ph/9712482, Phys. Rev. D58 (1998) 014014.

\bibitem{beauty}
H1 Collab. \PLB467 (1999) 156.\\
O. Deppe, (ZEUS Collab.), \photon99.\\ 
R. McNeil, (L3 Collab.), \photon99.

\bibitem{heravp}
H1 Collab., \PLB 415 (1997) 418-434 
and hep-ex/9812024, submitted to Eur. Phys. J. C.\\
Dorian Kcira, ZEUS Collab., \photon99.

\bibitem{dgsupp}
M. Drees, R. Godbole, \PRD 50 (1994) 3124.

\bibitem{eevp}
PLUTO collab., Phys. Lett. 142 (1984) 119.\\
P. Achard (L3 Collab.) \photon99.\\
M. Przybycien (OPAL Collab.) \photon99.

\bibitem{bfkl}
I.~Balitsky and L.N.~Lipatov, Sov. J. Nucl. Phys. 28 (1978) 822.

\end{thebibliography}
